\documentstyle[12pt]{article}
\title{
\begin{flushright}
{\normalsize Yaroslavl State University\\
             Preprint YARU-HE-97/06\\
             hep-ph/9709268} \\[20mm]
\end{flushright}
\Large \bf Creation of photons and electron - positron pairs 
by a neutrino in a strong magnetic field}
 
\author{A.A.~Gvozdev$^{1)}$, A.V.~Kuznetsov$^{1)}$, N.V.~Mikheev$^{1)}$, \\
L.A.~Vassilevskaya$^{1,2)}$\\
{\it 1) Yaroslavl State P.G.~Demidov University, Yaroslavl, Russia}\\
{\it 2) Moscow State M.V.~Lomonosov University, Moscow, Russia}}

\date{}

\begin{document}

\maketitle

\vspace{15mm}

\begin{abstract}
The processes of the photon and electron--positron pair production by a 
neutrino 
propagating in a strong magnetic field are investigated in the framework 
of the Standard Model. The process probabilities and the mean values 
of the neutrino energy and momentum loss are calculated. 
Possible astrophysical manifestations of the processes considered 
are briefly analysed.
\end{abstract}

\vspace{20mm}

\begin{center}
{\it Talk given at the 9th International School \\ Particles and Cosmology
\\ Baksan Valley, Kabardino Balkaria, Russia, April 15-22, 1997}
\end{center}

\newpage

\section{Introduction}

Neutrino physics in media is a vigorously growing
and prospective line of in\-ves\-ti\-ga\-ti\-ons at the junction of 
particle physics, astrophysics and cosmology.
Matter is usually 
considered as the medium. However, the strong magnetic field
can also play the role of an active medium. In the recent times an 
interesting effect of significant enhancement of the probability of the 
neutrino radiative decay $\nu_i \to \nu_j \gamma$ by a magnetic field 
(magnetic catalysis) was discovered in the frame of the Standard Model 
with lepton mixing~\cite{GMV,GMV96} (the effect was also confirmed in 
Refs.~\cite{Skob}). 

In the papers~\cite{GMV,GMV96} the massive neutrino radiative decay 
$\nu_i \rightarrow \nu_j + \gamma$ 
\footnote{Here $i$ and $j$ enumerate the neutrino mass eigenstates but not 
the neutrinos with definite flavors.} 
was studied in external electromagnetic fields of various configurations,
in particular, in the strong magnetic field. 
It was shown that a field-induced amplitude was not suppressed by the 
smallness of neutrino masses and did not vanish even in the case
of the massless neutrino as opposed to the vacuum amplitude.
The decay probability of the neutrino with energy $E_\nu < 2 m_e$
was calculated under the assumption that 
photon dispersion relation was close to the vacuum one, $q^2 = 0$.

However, the photon dispersion in the strong magnetic field
differs significantly from the vacuum dispersion 
with increasing of the photon energy~\cite{Ad}, so 
the on-shell photon  4-momentum can appear as the space-like  
with sufficiently large value of $q^2$, 
($\vert q^2 \vert \gg m^2_\nu$). 
In this case the phase space for the neutrino transition 
$\nu_i \rightarrow \nu_j + \gamma$ with $m_i < m_j$ is opened also. 
It means that the decay probability of ultrarelativistic neutrino 
becomes insensitive
to the neutrino mass spectrum due to the photon dispersion relation
in the strong magnetic field. This phenomenon results in a strong suppression
($ \sim m^2_\nu / E^2_\nu$) of the neutrino transition with flavour
violation, so a diagonal process 
$\nu_l \rightarrow \nu_l + \gamma$ ($l = e, \mu, \tau$)
is realized only. Thus, this diagonal radiative neutrino transition
does not contain 
uncertainties associated with a possible mixing in the lepton sector 
of the Standard Model, and can lead to observable physical effects in 
the strong magnetic fields.

Another decay channel also exists, $\nu_i \to \nu_j e^- e^+$, which is
forbidden in vacuum when $m_i < m_j + 2m_e$. 
However, the kinematics of a charged particle 
in a magnetic field is that which allows to have 
a sufficiently large space--like total momentum for 
the electron--positron pair, and this process is possible even for very light 
neutrinos. It means that a flavor of the ultrarelativistic neutrino is also 
conserved in this transition in a magnetic field, to the terms of the order 
of $m^2_{\nu}/E^2_{\nu}$ regardless of the lepton mixing angles. 
Consequently, a question of neutrino 
mixing is not pertinent in this case and the process 
$\nu \to \nu e^- e^+$ can be considered in the frame of 
the Standard Model without lepton mixing.

Here we investigate the processes $\nu \to \nu \gamma$ and 
$\nu \to \nu e^- e^+$ for the high 
energy neutrino, $E_{\nu} \gg m_e$, in a strong constant magnetic field. 
We consider a magnetic field as the strong one 
if it is much greater than the known Schwinger value 
$B_e = m^2_e/e \simeq 4.41 \cdot 10^{13} G$. 
These processes could be of importance in astrophysical applications, 
e.g. in an 
analysis of cataclysms like a supernova explosion or a coalescence of 
neutron stars, where the strong magnetic fields can exist, and where 
neutrino processes play the central physical role. 
In the present view, the magnetic field strength inside astrophysical 
objects in principle could be as high as 
$10^{16} - 10^{18} G$, both for toroidal~\cite{Lipu} and for
poloidal~\cite{Bocq} fields. The magnetic fields of the order of
$10^{12} - 10^{13} \, G$ have been observed at the surface of pulsars are 
the so-called `old' magnetic fields and they do not provide a determination 
of the field strength at a moment of the cataclysm.

Here we calculate the probabilities of the processes and the mean 
values of the neutrino energy and momentum loss through the production of 
electron--positron pairs and photons. 

If the momentum transferred is relatively small, $|q^2| \ll m^2_W$
\footnote{As the analysis shows, it corresponds in this case to the 
neutrino energy $E \ll m^3_W/e B$.}, the 
weak interaction of neutrinos with electrons could be described in the local 
limit by the effective Lagrangian of the form

\begin{equation}
{\cal L} \, = \, \frac{G_F}{\sqrt 2} 
\big [ \bar e \gamma_{\alpha} (g_V - g_A \gamma_5) e \big ] \,
\big [ \bar \nu \gamma^{\alpha} (1 - \gamma_5) \nu \big ] ,\quad
g_V = \pm {1 \over 2} + 2 sin^2 \theta_W \, , \quad g_A = \pm {1 \over 2}.
\label{eq:L}
\end{equation}

\noindent Here the upper signs correspond to the electron neutrino 
($\nu = \nu_e$) when both neutral and charged current interaction takes part 
in a process. The lower signs correspond to $\mu$ and $\tau$ neutrinos 
($\nu = \nu_{\mu}, \nu_{\tau}$), when the neutral current interaction 
is only presented in the Lagrangian~(\ref{eq:L}). 

The strong magnetic field is the only exotic we use. 

\section{The probability of the process $\nu \to \nu \gamma$}

The field-induced $\nu\nu\gamma$-vertex
can be calculated using an effective four-fermion
weak interaction~(\ref{eq:L}) 
of the left neutrino with the electron only, because
the electron is the most sensitive fermion to the external field.
By this means, the diagrams describing this process are reduced to an 
effective diagram with the electron in the loop 


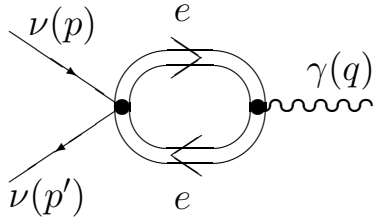
\begin{minipage}[t]{100mm}

\unitlength=1.00mm
\special{em:linewidth 0.4pt}
\linethickness{0.4pt}

\begin{picture}(60.00,35.00)(-40.00,13.00)

\put(35.00,32.50){\oval(20.00,15.00)[]}
\put(35.00,32.50){\oval(16.00,11.00)[]}
\put(26.00,32.50){\circle*{2.00}}
\put(44.00,32.50){\circle*{2.00}}

\linethickness{0.8pt}

\put(11.00,42.50){\vector(3,-2){9.00}}
\put(26.00,32.50){\line(-3,2){6.00}}
\put(26.00,32.50){\vector(-3,-2){9.00}}
\put(17.00,26.50){\line(-3,-2){6.00}}

\put(36.50,39.00){\line(-3,2){4.01}}
\put(36.50,39.00){\line(-3,-2){4.01}}

\put(32.50,26.00){\line(3,2){4.01}}
\put(32.50,26.00){\line(3,-2){4.01}}

\put(18.00,42.00){\makebox(0,0)[cb]{\large $\nu(p)$}}
\put(16.00,23.00){\makebox(0,0)[ct]{\large $\nu(p')$}}
\put(34.00,45.00){\makebox(0,0)[cc]{\large $e$}}
\put(34.00,20.00){\makebox(0,0)[cc]{\large $e$}}
\put(55.00,36.00){\makebox(0,0)[cb]{\large $\gamma(q)$}}


\def\photonatomright{\begin{picture}(3,1.5)(0,0)
                                \put(0,-0.75){\tencircw \symbol{2}}
                                \put(1.5,-0.75){\tencircw \symbol{1}}
                                \put(1.5,0.75){\tencircw \symbol{3}}
                                \put(3,0.75){\tencircw \symbol{0}}
                      \end{picture}
                     }
\def\photonrighthalf{\begin{picture}(30,1.5)(0,0)
                     \multiput(0,0)(3,0){5}{\photonatomright}
                  \end{picture}
                 }

\put(44.00,32.50){\photonrighthalf}

\end{picture}

\end{minipage}


\noindent where the double lines imply that the influence of the 
external field in the propagators is taken into
account exactly.

The calculation technique for the loop diagram of this type was 
described in detail in the paper~\cite{GMV96}.
We note that this effective $\nu\nu\gamma$ amplitude 
is enhanced substantially in the vicinity of the, so called, 
photon cyclotronic frequencies. The same phenomenon in the field-induced 
vacuum polarization is known as the cyclotronic resonance~\cite{Sh2}.

As was first shown by Adler~\cite{Ad}, two eigenmodes of the photon 
propagation with polarization vectors

\begin{eqnarray}
\varepsilon _{\mu}^{(\parallel)} & = & 
\frac{ (q \varphi)_{\mu} }{ \sqrt{ 
 q^2_\perp  } } ; \; \; \; \; \;
\varepsilon _{\mu}^{(\perp)} = \frac{ (q \tilde
\varphi)_{\mu} }{ \sqrt{ q^2_\parallel } }
\label{eq:EP}
\end{eqnarray}

\noindent are realized in the magnetic field, the, so-called, 
parallel ($\parallel$) and perpendicular ($\perp$) polarizations 
(Adler's notations).
Here $ \varphi_{\alpha \beta} = F_{\alpha \beta} / B$ and 
${\tilde \varphi}_{\alpha \beta} = \frac{1}{2} \varepsilon_{\alpha \beta
\mu \nu} \varphi_{\mu \nu} \; $ are the dimensionless tensor and dual
tensor of the external magnetic field with the strength $\vec B = (0,0,B)$;
$q^2_{\parallel}  =  ( q \tilde \varphi \tilde \varphi q ) =
q_\alpha \tilde \varphi_{\alpha\beta} \tilde \varphi_{\beta\mu} q_\mu 
= q^2_0 - q^2_3$,
$q^2_{\perp}  =  ( q \varphi \varphi q ) = q^2_1 + q^2_2$. 

It is of interest for some astrophysical applications
the case of relatively high neutrino energy 
$ E_\nu \simeq 10 - 20 MeV \gg m_e$ 
and strong magnetic field $e B > E^2_\nu$. 
As the analysis of the photon dispersion in a strong magnetic field shows, 
a region of the cyclotronic resonance $q^2_\parallel \sim 4 m_e^2$ 
in the phase space of the final photon, 
corresponding to the ground Landau level of virtual electrons dominates in the 
process $\nu \to \nu \gamma$. 
It is particularly remarkable that the $\perp$ photon mode only acquires 
a large space-like 4-momentum in the vicinity of the resonance. The 
corresponding amplitude of the process contains the enhancement due to 
the square-root singularity when $q^2_\parallel \to 4 m_e^2$. 
It means that taking account of the many-loop radiative corrections is needed. 
A detailed description of this procedure will be published elsewhere. 
A general expression of the probability of the process 
$\nu \to \nu \gamma^{(\perp)}$ has a rather comlicated form. 
Here we present the result of our calculation in the limit $e B \gg 
E^2_\nu \sin^2 \theta$: 

\begin{equation} 
W^{(\gamma)} \simeq \frac{\alpha G^2_F}{8 \pi^2 }(g^2_V + g^2_A) 
e^2 B^2 E \sin^2{\theta}.
\label{eq:w1}
\end{equation}

\noindent Here $E$ is the initial neutrino energy, 
$\theta$ is the angle between the vectors 
of the magnetic field strength ${\vec B}$ and the momentum of
the initial neutrino ${\vec p}$. The dispersion relation for the 
$\parallel$ photon mode is close to the vacuum one, $q^2 = 0$, and it gives 
a negligibly small contribution to the probability in the limit considered.

The probability of the process $\nu \to \nu \gamma^{(\perp)}$ is also 
non-zero above the threshold point of the $e^- e^+$-pair creation, 
$q^2_\parallel > 4 m_e^2$, due to an imaginary part of the amplitude. 
However, another channel $\nu \to \nu e^- e^+$ dominates in this region. 

\section{The probability of the process $\nu \to \nu e^- e^+$}

An amplitude of the process $\nu \to \nu e^- e^+$ could be immediately 
obtained from the Lagrangian~(\ref{eq:L}) where the known solutions of the 
Dirac equation in a magnetic field should be used. 


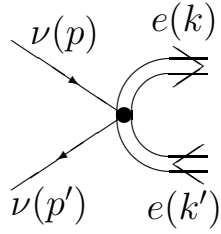
\begin{minipage}[t]{100mm}

\unitlength=1.00mm
\special{em:linewidth 0.4pt}
\linethickness{0.4pt}

\begin{picture}(60.00,35.00)(-40.00,13.00)

\put(37.00,32.50){\oval(24.00,15.00)[l]}
\put(37.00,32.50){\oval(20.00,11.00)[l]}
\put(26.00,32.50){\circle*{2.00}}

\linethickness{0.8pt}

\put(11.00,42.50){\vector(3,-2){9.00}}
\put(26.00,32.50){\line(-3,2){6.00}}
\put(26.00,32.50){\vector(-3,-2){9.00}}
\put(17.00,26.50){\line(-3,-2){6.00}}

\put(36.50,39.00){\line(-3,2){4.01}}
\put(36.50,39.00){\line(-3,-2){4.01}}

\put(32.50,26.00){\line(3,2){4.01}}
\put(32.50,26.00){\line(3,-2){4.01}}

\put(18.00,42.00){\makebox(0,0)[cb]{\large $\nu(p)$}}
\put(16.00,23.00){\makebox(0,0)[ct]{\large $\nu(p')$}}
\put(34.00,45.00){\makebox(0,0)[cc]{\large $e(k)$}}
\put(34.00,20.00){\makebox(0,0)[cc]{\large $e(k')$}}

\end{picture}

\end{minipage}


We present here the results of our calculations of the probability 
in two limiting cases which 
have a clear physical meaning: $e B \gg E^2 \sin^2 \theta$ and $e B \ll E^2 \sin^2 \theta$. 

In the case when the field strength $B$ appears to be 
the largest physical parameter, the electron and the positron could be born 
only in the states corresponding to the lowest Landau level. 
Integrating over the phase space, one obtains the following 
expression for the probability in the limit $e B \gg E^2 \sin^2 \theta$

\begin{equation}
W^{(e e)} = \frac{G_F^2 (g_V^2 + g_A^2)}{16 \pi^3} \, 
e B E^3 sin^4 \theta .
\label{eq:w2}
\end{equation}

In another limiting case, $e B \ll E^2 \sin^2 \theta$, when a great number of 
the Landau levels could be excited, our result for the process probability
can also be represented in a simple form

\begin{equation}
E W^{(e e)} \, = \, \frac{G_F^2 (g_V^2 + g_A^2)}{27 \pi^3} \, 
m^6 \chi^2 \, (ln \chi - \frac{1}{2} ln 3 - \gamma_E - \frac{29}{24}) ,
\label{eq:w3}
\end{equation}

\noindent  
where $\chi = e (pFFp)^{1/2}/m^3 = e B E sin \theta/m^3$ 
is the field dynamical parameter, $\gamma_E = 0.577 \dots$ is the Euler 
constant. 

\section{The neutrino energy and momentum losses}

It should be noted that a practical significance of these processes for 
astrophysics could be in the mean values of the neutrino energy and momentum 
losses rather than in the process probabilities.
These mean values could be found from the four-vector 

\begin{equation}
Q^\alpha \, = \, E \int d W q^\alpha \, = \, ({\cal I}, \vec {\cal F}) E.
\label{eq:Q}
\end{equation}

\noindent Its zero component is connected with the mean neutrino energy 
loss in a unit time, ${\cal I} = dE/dt$.
The space components of the four-vector~(\ref{eq:Q}) are connected 
similarly with the neutrino momentum 
loss in unit time, $\vec {\cal F} = d\vec p/dt$.
Here we present the expressions for $Q^{\alpha}$ in two limiting cases 
used above for the probability:

i) $e B \gg E^2 \sin^2 \theta$, for both processes

\begin{eqnarray}
{\cal I} & = & E W \, C_1 \, 
\left(1 + \frac{2 g_V g_A}{g_V^2 + g_A^2} \cos\theta\right),
\label{eq:I} \\[2mm]
{\cal F}_z & =  & E W \, C_1 \, 
\left(\cos\theta  + \frac{2 g_V g_A}{g_V^2 + g_A^2}\right), \quad
{\cal F}_\perp \, =  \, E W \, C_2 \, \sin\theta,
\label{eq:F} 
\end{eqnarray}

\noindent 
where the $z$ axis is directed along the field, the vector 
$\vec {\cal F}_\perp$, transverse to the field, lies in the plane of 
the vectors $\vec B$ and $\vec p$,

$$C_1^{(\gamma)} = \frac{1}{4},
\quad
C_2^{(\gamma)} = \frac{1}{2},
\quad
C_1^{(ee)} = \frac{1}{3},
\quad
C_2^{(ee)} = 1.$$

ii) $e B \ll E^2 \sin^2 \theta$, for the process $\nu \to \nu e e$

\begin{equation}
Q^\alpha \, \simeq \, \frac{7}{16} E W^{(ee)} \, p^\alpha,
\label{eq:Q2}
\end{equation}

\noindent 
where the probability should be taken from Eq.~(\ref{eq:w3}).
All expressions for the process $\nu \to \nu e^- e^+$
are also applicable for the process with antineutrino 
$\tilde\nu \to \tilde\nu e^- e^+$ due to the $CP$--invariance of the weak 
interaction.

\section{An applicability of results in a presence of hot dense plasma}

Let us note that our results are valid in the presence of plasma with the 
electron density $n \sim 10^{33} - 10^{34} cm^{-3}, 
\, (n = n_{e^-} - n_{e^+})$. This is due to a peculiarity of the statistics 
of the relativistic electron gas in a magnetic field~\cite{Lan}.
A dependence of the density of the relativistic electron-positron gas on the 
chemical potential $\mu$ and the temperature $T$ can be written, in view of 
the degeneration over the transversal momentum, as the following sum over the 
Landau levels

\begin{eqnarray}
n & = & 
 n_{e^-} - n_{e^+} =  
\frac{e B}{2 \pi^2} \int\limits_0^\infty d p \, \bigg \lbrace
\left(\exp{\left({p - \mu \over T}\right)} + 1 \right)^{-1} +
\nonumber\\[5mm]
& + & 2 \sum_{k=1}^\infty
\left(\exp{\left({\sqrt{p^2 + 2 k e B} - \mu \over T}\right)} + 1 \right)^{-1} 
\, - \, (\mu \to - \mu)\bigg \rbrace .
\label{eq:n1}
\end{eqnarray}

In the strong magnetic field, $\sqrt{e B} - \mu \gg T$, when the ground Landau 
level is only occupied, a dependence on the temperature in Eq.~(\ref{eq:n1}) 
disappears. The chemical potential as a function of the density becomes

\begin{equation}
\mu  =  \frac{2 \pi^2 n}{e B} \,\simeq \, 
 0.26 ~\mbox{MeV} 
\left (\frac{n}{10^{33}~\mbox{cm}^{-3}} \right )
\left (\frac{10^{17}~\mbox{G}}{B} \right ) .
\label{eq:mu}
\end{equation}

\noindent  
It is significantly less for the same density than in the case without field

\begin{equation}
\mu_0 =  (3 \pi^2 n)^{1/3} \,\simeq \, 
 6.1 ~\mbox{MeV} 
\left (\frac{n}{10^{33}~\mbox{cm}^{-3}} \right )^{1/3}.
\label{eq:mu0}
\end{equation}

\noindent  
As the analysis shows, 
the suppressing statistical factors in integrating over the phase 
space do not arise at the conditions
$ B > 10^{15} \;\mbox{G} \;(T/3~\mbox{MeV})^2$ and

\noindent 
$ B > 5\cdot10^{15} \;\mbox{G} \;
(n/10^{33}~\mbox{cm}^{-3})^{2/3}.$

In another limiting case of very high temperature, $T \gg \sqrt{eB},\, \mu$, 
an influence of a medium leads to the constant statistical factors of 1/2 both 
for an electron and for a positron. This reduces the process probability 
in 4 times.

\section{Possible astrophysical consequences}

To illustrate the formulae obtained we consider 
the astrophysical process of a birth of the magnetized neutron star, 
pulsar, for example, in a supernova explosion. 
Let us suppose that in the cataclysm a very strong magnetic field of the order 
of $10^{16} - 10^{18} G$~\cite{Lipu, Bocq} arises in some reason in the 
vicinity of a neutrinosphere. The electron density in this region will 
be considered to be not too high, so 
a creation of the $e^- e^+$ pairs is not suppressed by statistical factors. 
In this case the neutrino propagating through the 
magnetic field will loose the energy and the momentum in accordance with 
our formulas. A part of the total energy lost by neutrinos in the field 
due to the process of the $e^- e^+$ pair creation
could be estimated from Eq.~(\ref{eq:I}):

\begin{equation}
\frac{\Delta {\cal E}^{(ee)}}{{\cal E}_{tot}} 
  \sim  0.6 \cdot 10^{-2}\, 
\left (\frac{B}{10^{17}~\mbox{G}} \right )\,
\left (\frac{\bar E}{10 ~\mbox{MeV}}\right )^3\,
\left (\frac{ \Delta \ell}{10 ~\mbox{km}}\right ), 
\label{eq:DE1}
\end{equation}

\noindent here $\Delta \ell$ is a characteristic size of the region 
where the field strength varies in\-signi\-fi\-cant\-ly, 
${\cal E}_{tot}$ is the total energy carried off by
neutrinos in a supernova explosion, 
$\bar E$ is the neutrino energy averaged over the neutrino spectrum.
Here we take the energy scales which are believed to be typical for 
supernova explosions~\cite{Imsh}.

An asymmetry of outgoing neutrinos is another interesting manifestation

\begin{equation}
A \; = \; \frac{|\sum_i {\bf p}_i|}{\sum_i |{\bf p}_i|}.
\label{eq:A1}
\end{equation}

\noindent In the same limit of the strong field we obtain 

\begin{equation}
A^{(ee)}  \sim  2 \cdot 10^{-3}\, 
\left (\frac{B}{10^{17}~\mbox{G}} \right )
\left (\frac{\bar E}{10~\mbox{MeV}}\right )^3 
\left (\frac{ \Delta \ell}{10~\mbox{km}}\right ).
\label{eq:A2}
\end{equation}

\noindent  
One can see from Eqs.~(\ref{eq:DE1}), (\ref{eq:A2}) that the effect could 
manifest itself at a level of about percent. In principle, it could be 
essential in a detailed theoretical description of the process of 
supernova explosion. 
For the process $\nu \to \nu \gamma$ one obtains 

\begin{equation}
A^{(\gamma)}  \sim  2 \pi \alpha \frac{e B}{E^2} A^{(ee)}. 
\label{eq:A3}
\end{equation}

\noindent  
It is seen that two processes considered could be comparable for some values 
of physical parameters. 

In another limiting case $e B \ll E^2 \sin^2 \theta$ we obtain from 
Eq.~(\ref{eq:Q2}) 

\begin{equation}
\frac{\Delta {\cal E}^{(ee)}}{{\cal E}_{tot}} 
  \sim  10^{-6}\, 
\left (\frac{B}{10^{15}~\mbox{G}} \right )^2\,
\left (\frac{\bar E}{20 ~\mbox{MeV}}\right )\,
\left (\frac{ \Delta \ell}{10 ~\mbox{km}}\right ) \, 
\left [ 4.7 + ln \left (\frac{B}{10^{15}~\mbox{G}} \,
\frac{\bar E}{20 ~\mbox{MeV}}\right ) \right ].
\label{eq:DE2}
\end{equation}

Let us note that an origin of the asymmetry of the neutrino momentum loss 
with respect to the magnetic field direction is a manifestation 
of the parity violation in weak interaction, because the ${\cal F}_z$ value 
contains the term proportional to the product of 
the constants $g_V$ and $g_A$. This asymmetry could result in the recoil  
``kick'' velocity of a rest of the cataclysm. 

\section{Conclusions}

\noindent $\bullet$  
If the physical parameters would have the above-mentioned values, 
the effect could 
manifest itself at a percent level. It could be 
essential in a detailed theoretical description of the process of a 
supernova explosion or a coalescence of neutron stars.

\noindent $\bullet$  
An origin of the asymmetry of the neutrino momentum loss 
with respect to the magnetic field direction is a manifestation 
of the parity violation in weak interaction
(proportional to $g_V g_A$). 

\noindent $\bullet$  
This asymmetry results in the recoil 
``kick'' velocity of a rest of the cataclysm. For the parameters used, 
it would provide a ``kick'' velocity 
of order 150 km/s for a pulsar with mass of order of the solar mass. 

\noindent $\bullet$  
Produced $e^- e^+$ pairs and $\gamma$-quanta are captured 
by a strong magnetic field and propagate along the 
field~\cite{Sh2}. Thus a mechanism of
significant power ``pumping" of polar caps of a magnetized 
remnant could take place.

\bigskip

\noindent {\bf Acknowledgements}  
The authors are grateful to V.M.~Lipunov, V.N.~Lukash, V.G. Kurt, 
K.A.~Postnov, and M.E.~Prokhorov for useful discussions. 
A.K. expresses his deep gratitude to the Organizing Committee of the 
9 International Baksan School ``Particles and Cosmology'' for the possibility 
to participate in this School. This work is supported in part by the
International Soros Science Education Program under the Grants 
N~d97-872 (A.K.) and N~d97-900 (A.G.).

\end{document}